\documentclass[preprint]{aastex}

\pdfoutput=1

\newcommand{\etal}{et al.\ }
\newcommand{\eg}{e.g. }

\usepackage{epstopdf}
\usepackage{graphicx}
\usepackage{color}
\usepackage{epsfig}

\def\Fig#1{Figure~\ref{fig:#1}}
\def\se#1{\S\ref{sec:#1}}

\shorttitle{Study of Isochrone Consistency}
\shortauthors{Hills \etal}

\begin{document}

\title{BAYESIAN INVESTIGATION OF ISOCHRONE CONSISTENCY USING THE OLD OPEN
CLUSTER NGC 188\footnote{This is paper 64 of the WIYN Open Cluster Study (WOCS).}}

\author{Shane Hills}
\affil{Department of Physics, Engineering Physics and Astronomy, Queen's
University, Kingston, ON K7L 3N6 Canada}
\email{shane.hills@queensu.ca}

\author{Ted von Hippel}
\affil{Department of Physical Sciences, Embry-Riddle Aeronautical University, 
Daytona Beach, FL 32114, USA}
\email{ted.vonhippel@erau.edu}

\author{St\'{e}phane Courteau}
\affil{Department of Physics, Engineering Physics and Astronomy, Queen's
University, Kingston, ON K7L 3N6 Canada}
\email{courteau@astro.queensu.ca}

\and
\author{Aaron M.\ Geller}
\affil{Center for Interdisciplinary Exploration and Research in
Astrophysics (CIERA) and Department of Physics and Astronomy, Northwestern
University, 2145 Sheridan Rd, Evanston, IL 60208, USA}
\affil{Department of Astronomy and Astrophysics, University of Chicago,
5640 S. Ellis Avenue, Chicago, IL 60637, USA}
\email{a-geller@northwestern.edu}

\begin{abstract}

This paper provides a detailed comparison of the differences in parameters
derived for a star cluster from its color-magnitude diagrams depending on
the filters and models used.  We examine the consistency and reliability of
fitting three widely-used stellar evolution models to fifteen combinations
of optical and near-IR photometry for the old open cluster NGC 188.  The
optical filter response curves match those of the theoretical systems and
are thus not the source of fit inconsistencies.  NGC 188 is ideally suited to
the present study thanks to a wide variety of high-quality photometry and
available proper motions and radial velocities which enable us to remove
non-cluster members and many binaries.  Our Bayesian fitting technique
yields inferred values of age, metallicity, distance modulus, and
absorption as a function of the photometric band combinations and stellar
models.  We show that the historically-favored three band combinations of
$UBV$ and $VRI$ can be meaningfully inconsistent with each other and with
longer baseline datasets such as $UBVRIJHK_S$.  Differences among model
sets can also be substantial.  For instance, fitting Yi \etal (2001) and
Dotter \etal (2008) models to $UBVRIJHK_S$ photometry for NGC 188 yields
the following cluster parameters: age = \{5.78 $\pm$ 0.03, 6.45 $\pm$
0.04\} Gyr, [Fe/H] = \{+0.125 $\pm$ 0.003, $-$0.077 $\pm$ 0.003\} dex,
$(m-M)_{\rm V}$ = \{11.441 $\pm$ 0.007, 11.525 $\pm$ 0.005\} mag, and
$A_{\rm V}$ = \{0.162 $\pm$ 0.003, 0.236 $\pm$ 0.003\} mag, respectively.
Within the formal fitting errors, these two fits are substantially and
statistically different.  Such differences amongst fits using different
filters and models are a cautionary tale regarding our current ability to
fit star cluster color-magnitude diagrams.  Additional modeling of this
kind, with more models and star clusters, and future GAIA parallaxes are
critical for isolating and quantifying the most relevant uncertainties in
stellar evolutionary models.  \end{abstract}


\section{Introduction}

Stellar evolution is a mature field with numerous successes, perhaps the
most important of which is the ability to determine star cluster ages.
These ages are the basis for our understanding of the star formation
histories of the Milky Way and other galaxies and provide a cornerstone of
modern astrophysics. Yet, the ages that we derive for star clusters suffer
from well-known observational and theoretical uncertainties (\eg Kurucz
2002; Asplund \etal 2009; Pereira \etal 2013), as well as difficulties in
matching observations to theory (\eg Flower 1996; von Hippel \etal 2002;
Dotter \etal 2008). We focus on these data-model comparisons, which have
historically been exacerbated by subjective fitting techniques and an
unknown sensitivity on filter choice.  Generally, researchers adjust a
handful of model parameters, testing for their impact in color-magnitude
diagrams (CMDs), until a good match is found.  This approach is subjective
as different research groups matching the same multi-band data set to the
same models might not derive matching cluster parameters.  Furthermore,
studies of the same cluster by different groups may differ in their
choice of photometric bands.  Yet, the sensitivity of cluster parameters
to different filter combinations remains poorly constrained, though past
studies (\eg Figure 10 of Sarajedini \etal 1999; Grocholski \& Sarajedini 2003)
indicate that stellar modeling results do depend on the choice of photometric bands.

In this paper, we refine the data-model interface by employing an
objective Bayesian fitting technique and using it to study the
sensitivity of fits between stellar models and common subsets of $UBVRIJHK_S$
photometry.  We focus our investigation on the old open cluster NGC 188
because of the available, extensive, high-quality photometry and the
high-quality proper-motion and radial-velocity data, which enable us to
remove most non-cluster members and binaries from the CMD.  By focusing
on one cluster, we can perform an extensive CMD analysis with many filter
combinations and three stellar evolution codes, yet our results are
necessarily limited to clusters with parameters similar to NGC 188.
Needless to say, this approach ought to motivate similar studies with
a broad suite of stellar evolution models and wide range of stellar clusters.

We first describe the photometry, radial velocities, and proper motions
for NGC 188 in section \se{data}. We then present the Bayesian statistical
technique used for our analysis in section \se{method}, and apply it to
different photometry band combinations in section \se{results}.  Our
results and conclusions, with a view to future investigations and an
extension of this analysis to other clusters, are presented in sections
\se{discussion} and \se{conclusion}, respectively.

\section{Observations}
\label{sec:data}

For the observational data of NGC 188, we rely on the homogenized $UBVRI$
photometry of Stetson, McClure, \& VandenBerg (2004), which were derived
from nearly a dozen independent observational studies, combined with
cluster membership probabilities from four proper-motion studies.  The
sheer number of previous photometric and astrometric studies of NGC 188
over the last fifty years (Sandage 1962; Sharov 1965; Cannon 1968; Eggen \&
Sandage 1969; Upgren \etal 1972; McClure \& Twarog 1977; Caputo \etal 1990;
Dinescu \etal 1996; von Hippel \& Sarajedini 1998; Sarajedini \etal 1999;
Platais \etal 2003) demonstrate the considerable interest in this cluster
largely because it is well-populated, relatively unreddened, and one of the
oldest known open clusters.  We supplement the optical photometry with
2MASS $JHK_S$ photometry from Skrutskie \etal (2006).

Along with the photometry, we take advantage of the multi-epoch
radial-velocity data from Geller \etal (2008; 2009; and additional
observations from the continuation of their radial-velocity survey).  This
radial-velocity sample contains at least three observations for nearly
every solar-type star from $\sim$1.5 mag below the cluster turn-off up to
the tip of the giant branch ($\sim$0.95 to 1.15 M$_{\odot}$) within a
1\degr\ diameter region centered on NGC 188 (corresponding to a radius of
$\sim$17 pc or roughly 13 core radii), and spans a monitoring baseline of
more than a decade for many stars.  The radial-velocity data enable us to
further remove non-cluster members and many of the cluster binary stars in
what is otherwise a busy, field-star-contaminated CMD.  The radial-velocity
survey's ability to detect binaries depends primarily on the orbital
periods because of the monitoring baseline of the survey, and is also
sensitive to the binary mass ratios, eccentricities, and orbital
inclination (see Geller \etal 2012).  Monte Carlo completeness analysis
indicates that 63\% of the solar-type binaries with orbital periods $<10^4$
days (of all mass ratios, eccentricities, and inclinations) are detected in
this survey, whereas very few with longer periods are detected. Geller
\etal (2012) estimate the hard-binary versus soft-binary boundary for NGC
188 to be about $10^6$ days, so presumably some fraction of long-period
solar-type binaries remain undetected.  Yet, removing this large sample of
binaries is particularly useful for accurate cluster quantities because
such binaries can confuse the location of the main sequence turn-off and
substantially broaden the main sequence.  Our analysis technique (see
below) includes fitting binaries of all mass ratios (regardless of orbital
periods, eccentricities, etc.), yet our results are more accurate when we
can remove these binaries in advance.

\begin{figure}[!h]
\centering
\includegraphics[width=\textwidth]{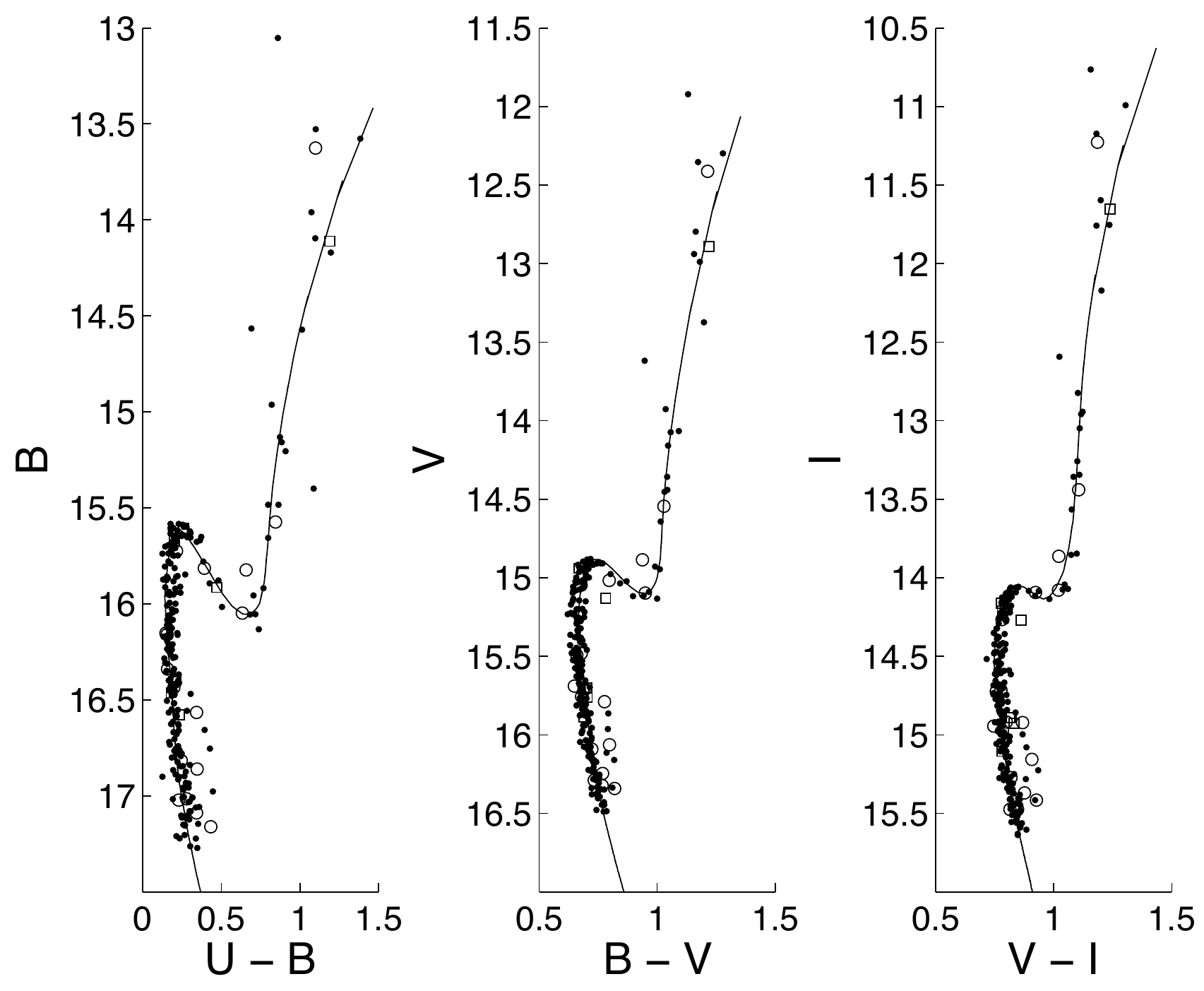}
\caption{CMDs for NGC 188 in a range of optical bands. The point types
correspond to the assigned cluster membership probability for each star
(filled circles have membership probabilities greater than 0.9, open
circles indicate between 0.7 and 0.9 probability, open squares show
probabilities less than 0.7). These probabilities result from the
arithmetic mean of membership probabilities from Stetson \etal (2004)
(based on proper motions) and Geller \etal (2008) (based on radial
velocities). The data are overplotted with the mean fitted isochrone (solid
line) found by BASE-9 in its $UBVRI$ photometry fit to Yi \etal (2001)
models (see section \ref{sec:results}).}
\label{fig:cmdfit}
\end{figure}

We obtain a well-populated CMD with 248 stars, each with full $UBVRIJHK_S$
photometry and a variety of cluster membership metrics, by combining the
optical photometry and proper-motion data from Stetson \etal (2004) with
the infrared photometry from Skrutskie \etal (2006), as well as the
radial-velocity measurements from Geller \etal (2012), including all
star-by-star uncertainties in every parameter.  \Fig{cmdfit} shows the
CMD for NGC 188.

We considered multiple ways to incorporate both the proper-motion and
radial-velocity cluster membership probabilities for this study, though
none are ideal. Simply multiplying these independent probabilities yields
very low membership probabilities (i.e. $\sim 20\%$ or less) for some
stars, potentially due to uncertainties in one set of data or the other.
This reduces the importance of those stars to the fit.  The alternative of
taking the greatest cluster membership value would neglect potentially
important information contained in whichever method was ignored.  We
explored a few other techniques of combining cluster memberships.
Fortunately, our tests with various datasets showed that the results were
insensitive to the specific techniques used to combine membership
probabilities.  As a result we employ a simple arithmetic mean of the
proper-motion and radial-velocity probabilities for our cluster membership
probabilities.

\section{Statistical Method}
\label{sec:method}

For the purposes of this study, we require an objective and precise
technique to fit stellar isochrones to cluster photometry.  The software
suite BASE-9\footnote{BASE-9 is freely available from the second author.}
({\bf B}ayesian {\bf A}nalysis of {\bf S}tellar {\bf E}volution with {\bf
9} Parameters) fits our requirements well.  (For a full discussion of the
method, software, and the capabilities of BASE-9, see von Hippel \etal
2006; DeGennaro \etal 2009; van Dyk \etal 2009; and Stein \etal 2013).)
BASE-9 compares stellar evolution models (listed below) to photometry in
any combination of photometric bands for which there are data and models.
Of particular benefit for our study, BASE-9 accounts for individual errors
for every data point, incorporates cluster membership probabilities from
proper motions or radial velocities, and cluster metallicity from
spectroscopic studies. As per above, we incorporated membership
probabilities from both proper motions and radial velocities using an
arithmetic mean.

BASE-9 uses a computational technique known as Markov chain Monte Carlo
(MCMC) to derive the Bayesian joint posterior probability distribution for
up to six parameter categories (cluster age, metallicity, helium content,
distance, and reddening, and optionally for white dwarf studies, a
parameterized initial-final mass relation) and brute-force numerical
integration for three parameter categories (stellar mass on the zero-age
main sequence, binarity, and cluster membership).  The last three of these
parameter categories include one parameter per star whereas the first six
parameter categories refer to the entire cluster.  This study includes no
white dwarfs and we therefore do not use the initial-final mass relation.
Additionally, among the isochrone sets we employ, there is a fixed
relationship between metallicity and helium content.  We are also not
concerned with the individual stellar masses in this study.  BASE-9
marginalizes over the parameters that are not of direct interest to us,
yielding the four cluster-wide parameters (age, metallicity, distance,
absorption) pertinent to this work.

BASE-9 allows us to take advantage of prior information, where available,
to constrain parameters.  For this problem, the results are insensitive to
the exact choice of reasonable priors.  We chose priors with a Gaussian
shape in the logarithmic quanitities with mean values from the photometric
study of Sarajedini \etal (1999), specifically [Fe/H] = $-0.03$,
$(m-M)_{\rm V}$ = 11.44, and $A_{\rm V}$ = 0.3.  Yet, we set the
uncertainties on these priors to be broad enough that they would not
unreasonably constrain our fits, given our current knowledge of these
values, specifically $\sigma$([Fe/H]) = 0.3, $\sigma((m-$M)$_{\rm V}$) =
0.3, and $\sigma(A_{\rm V}$) = 0.1.  We performed sensitivity tests on
these priors and found that reasonable values for these prior standard
deviations yielded negligible differences compared to the variation caused
by different filter combinations.

We have adopted the stellar evolution models of Girardi \etal (2000), Yi
\etal (2001), and Dotter \etal (2008).  The Girardi \etal isochrones span
the age range of 63 Myr to nearly 18 Gyr from metal-free stars up to [Fe/H]
= +0.2.  The Yi \etal isochrones span 1 Myr to 20 Gyr from [Fe/H]=$-3.7$ to
nearly $+0.8$.  The Dotter \etal isochrones span 250 Myrs to 15 Gyrs over a
metallicity range of [Fe/H] = $-2.5$ to $+0.5$.  All of these parameter
ranges easily bracket NGC 188, which has near solar abundance and is
approximately 6 Gyrs old.

\section{Results}
\label{sec:results}

\subsection{Posterior Probability Distributions and Best-Fit Isochrones}

It is common practice to run MCMC routines such as BASE-9 (von Hippel \etal 2006)
long enough to collect 10,000 uncorrelated samples.  In many cases, if the
MCMC steps are too small, the excursions through parameter space are
correlated, yielding fewer independent samples.  In such a case, we then
keep only every n-th iteration, where n is set such that subsequent stored
iterations are uncorrelated.  While there is no well-defined number of
appropriate samples, 10,000 is usually more than sufficient to determine
the shape of every posterior distribution function, including those with
extended tails or multi-modality.  At the other extreme, the Central
Limit Theorem dictates that approximately 30 uncorrelated samples are
sufficient for a normal distribution (\eg Hogg \& Tanis 2005).  Before
running a particular dataset against a specific set of models, we do not
know if the posterior distributions will be Gaussian shaped or more
complex, so we take the conservative approach and initially assume complex
posterior distributions and run BASE-9 for 10,000 uncorrelated iterations.

\begin{figure}[!h]
\epsscale{1.0}
\plottwo{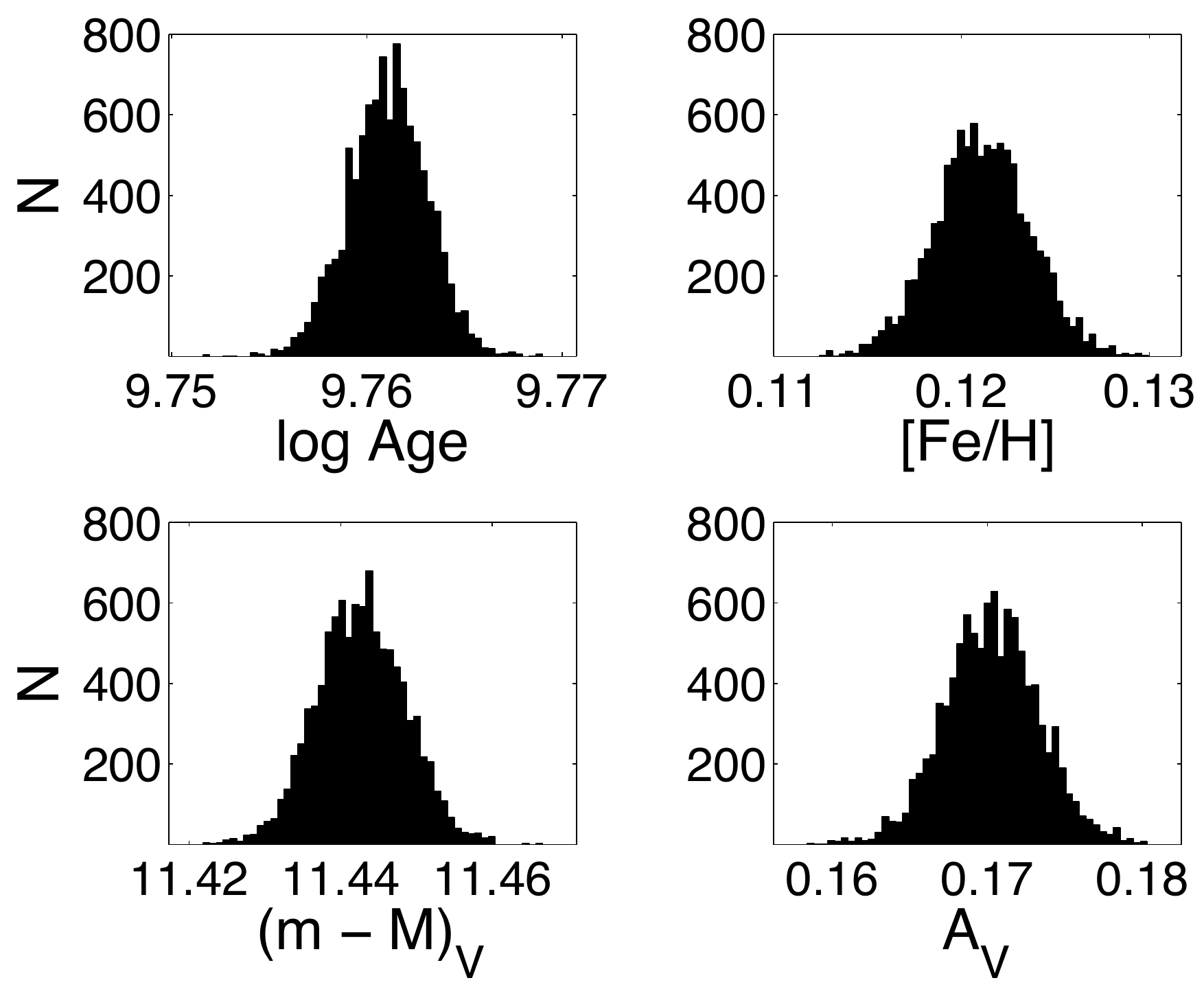}{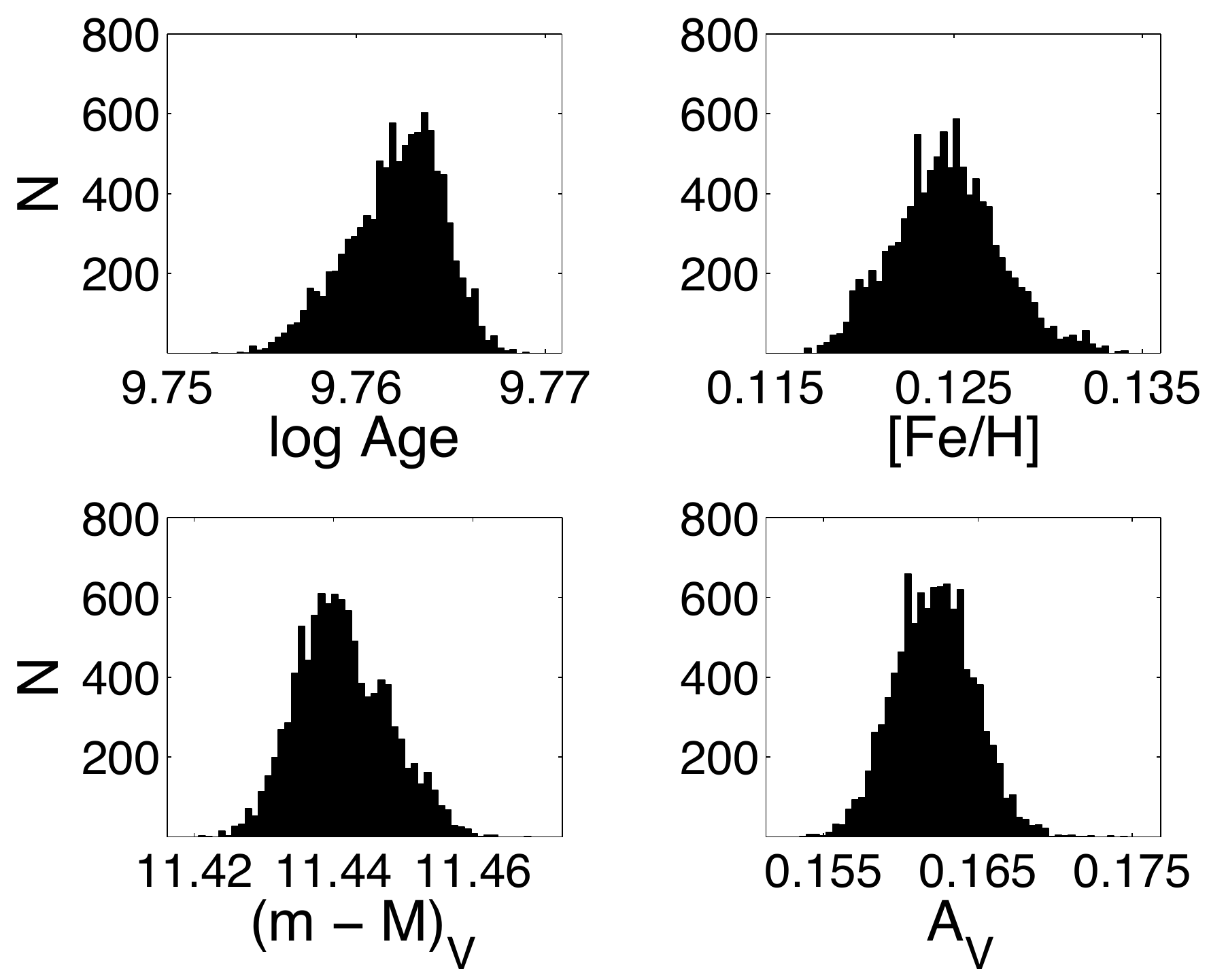}
\caption{{\bf (Left)} Posterior probability distributions for log age,
[Fe/H], $(m-M)_{\rm V}$, and $A_{\rm V}$ using $UBVRI$ photometry and Yi
\etal (2001) isochrones. {\bf (Right)} Posterior probability distributions
for these same cluster parameters based on $UBVRIJHK_S$ photometry and Yi
\etal (2001) isochrones.  The log age distribution is noticeably skewed,
which is also the case with the Dotter \etal (2008) models.}
\label{fig:paramhist}
\end{figure}

Each of the uncorrelated samples from BASE-9 is an allowable fit of a
particular family of stellar evolution models to the cluster data, given the
photometric errors, probabilities of membership, etc.  In order to obtain
the posterior probability distributions for the four cluster parameters of
interest (age, [Fe/H], ($m-M)_{\rm V}$, $A_{\rm V}$), we marginalize the
sampling history derived by BASE-9 by binning along the parameter axis of
interest. Many of these distributions are nearly Gaussian in shape, however
some are substantially non-Gaussian.  The root cause of the non-Gaussian
distributions is that stellar evolution is intrinsically non-linear, so
that, for example, Gaussian errors in photometry do not propagate as
Gaussian errors in cluster age.  One of the strengths of the Bayesian
technique is that it recovers the posterior parameter distributions, which
provide an informative indication of uncertainty that often cannot be
captured with a simple (frequentist) best-fit parameter with error bars.
\Fig{paramhist} presents posterior probability distributions for two
different filter combinations.

As an example of the fits derived by BASE-9, we overplot in \Fig{cmdfit} an
isochrone derived from the mean parameters for the Yi \etal (2001) stellar
evolution models fit to the $UBVRI$ dataset, the posterior distributions of
which are plotted in the four panels on the left in \Fig{paramhist}.  These
particular distributions are symmetric, so the mean and median cluster
parameters are essentially identical.  In some cases they are different,
yet typically with the precision of this technique and the quality of this
dataset there are no differences to the human eye between isochrones based
on the mean versus the median fit.  The Bayesian approach also reminds us
that there is no single best fit isochrone, but rather a range of
probabilistically acceptable isochrones.  It is the distribution of these
acceptable isochrones that forms the posterior distributions.  Any
overplotted isochrone is at best just a representative example drawn from
that distribution.  In fact, isochrones created from summary statistics
such as mean or median parameters may not be truly representative if the
distributions are substantially non-Gaussian because that simultaneous
combination of parameters may fit the data with low probability.

\subsection{Differences Among Commonly Used Filters}

\begin{deluxetable}{llrrrr}
\tablenum{1}
\tablehead{
\colhead{filters} & 
\colhead{prescription} & 
\colhead{$\Delta$(age)} & 
\colhead{$\Delta$([Fe/H])} & 
\colhead{$\Delta$(m-M)} & 
\colhead{$\Delta$($A_{\rm V}$)} \\ 
\colhead{} & 
\colhead{} & 
\colhead{Gyr} & 
\colhead{dex} & 
\colhead{mag} & 
\colhead{mag} }
\startdata
UB      & Bessell 2      & $-0.04$ (0.13) &   0.01  (0.012)  & $-0.03$ (0.014) &   0.06  (0.041) \\
UB      & Maiz-Apellaniz & $-0.38$ (0.14) & $-0.03$ (0.011)  & $-0.02$ (0.019) &   0.33  (0.040) \\
UBV     & Bessell 2      &   1.03  (0.12) &   0.03  (0.006)  & $-0.07$ (0.012) & $-0.13$ (0.013) \\
UBV     & Maiz-Apellaniz &   1.64  (0.06) &   0.05  (0.005)  & $-0.09$ (0.011) & $-0.14$ (0.009) \\
UBVRI   & Bessell 2      &   0.11  (0.06) &   0.01  (0.004)  & $-0.02$ (0.008) & $-0.03$ (0.005) \\
UBVRI   & Maiz-Apellaniz &   0.42  (0.05) &   0.05  (0.003)  & $-0.05$ (0.007) & $-0.04$ (0.005) \\
\enddata
\tablecomments{All values are relative to the first Bessell prescription.}
\end{deluxetable}

Before embarking on a detailed comparison of model fits as a function of
subsets of the adopted $UBVRIJHK_S$ filters, we take a brief look into
the subject of filter prescriptions.  Because observatories have
different versions of these filters and stellar isochrones do not include
all possible such filters, we sought to check the sensitivity of BASE-9
fits to a few filter prescriptions.  Specifically, we test three variations
on the $B$ filter from Bessell (1990) and Maiz-Apellaniz (2006) by using
BASE-9 to fit simulated $UB$, $UBV$, and $UBVRI$ photometry based on these
filter prescriptions.  These sensitivity tests are clearly not exhaustive.
Rather, they are meant to provide an estimate of the sensitivity isochrone
fits have to filter prescriptions.

We employed Girardi \etal models, available at
http://stev.oapd.inaf.it/cgi-bin/cmd (see Bressan \etal 2012), to generate
isochrones with Z = 0.0144, Y = 0.27387, [M/H] = $-$0.01, and age = 6.41
Gyr, along with either of the two $B$ filters of Bessell (1990) or the $B$
filter of Maiz-Apellaniz (2006).  These model parameters are almost
identical to one set of our NGC 188 fits (see next section).  To these
isochrones, we added a distance modulus of 11.38 and offset the absorption
for all bands according to Table 3 of Cardelli, Clayton, \& Mathis (1989)
for $A_{\rm V}$ = 0.17.  We then created approximately the same number of
simulated stars along this sequence as we have stars in our NGC 188
database and added appropriately-sized photometric errors.

We recovered the four cluster parameters for these simulated clusters with
BASE-9.  Table 1 presents a comparison among these fits, with offsets
between any particular fit and the fit to the first Bessell prescription
for that filter combination.  Uncertainties derived from combining in
quadrature the one standard deviation ranges for both posterior probability
distributions are listed in parentheses.   The differences between the
two Bessell fits in the $UB$ case are zero within the errors.  They are
statistically significant in the $UBVRI$ case, but small.  For the $UBV$
case, the differences are substantial and amount to 1.03 $\pm$ 0.12 Gyr,
though the other cluster parameters are more stable.  The differences between
the fit to the first Bessell $B$-filter prescription and that of Maiz-Apellaniz
is substantial and statistically significant in all three cases, particularly
for ages.  These differences arise from assuming that the $B$ filter used
in the simulated observations is the same as the $B$ filter used in the
fit, highlighting the importance of incorporating the correct filter
prescriptions whenever possible, and particularly when deriving absolute
cluster parameters.

While filter prescriptions fundamentally matter, we defer related studies
to a future paper given that the comparisons among filters presented in the
next section are {\it differential}.  We have a high-quality data set that
undoubtedly suffers small systematics (estimated by Stetson, McClure, \&
VandenBerg (2004) to be $\leq$ 0.02 mag for the optical data).  We will use
subsets of filters from this same data set repeatedly and compare these
data to the same models, looking for differences among fits as a function
of filter within a given stellar model set.

\subsection{Cluster Parameters as a Function of Selected Filters}

We fit three widely used stellar evolution models (Girardi \etal 2000; Yi
\etal 2001; Dotter \etal 2008) to fifteen combinations of optical/near-IR
photometry.  These three isochrone sets consistently rely on
Johnson-Cousins $UBVRI$ as defined by Bessell (1979, 1990).  We acknowledge
a filter mismatch with the Yi \etal and Girardi \etal models for which a
$K_S$ filter response is not available.  Because the $JHK$ age constraints
are not as reliable as those at optical wavelengths, due partly to the lack
of isochrone morphology information in these red bandpasses, we do not
report $JHK$ fits nor attempt a $K_S$ to $K$ transformation, which would
introduce additional uncertainty.

Our BASE-9 fits produced too many fits to present all the posterior
distributions.  Additionally, we require summary statistics in order to
compare among these models and filters.  Therefore, we adopt
box-and-whisker plots to provide both summary statistics and capture the
degree of non-Gaussianity in the distributions.  In box-and-whisker plots,
the central line delineates the median of the distribution and the box
edges indicate the 25th and 75th percentiles. The whiskers extend out to
the most extreme non-outliers, and outliers are plotted individually. A
data point is considered an outlier if it is smaller than $q_1 -
\frac{3}{2}(q_3 - q_1)$ or greater than $q_3 + \frac{3}{2}(q_3 - q_1)$,
where $q_1$ and $q_3$ are the 25th and 75th percentiles, respectively.

\begin{figure}[htp]
  \centering
  \begin{tabular}{cc}
    \includegraphics[angle=0,width=85mm]{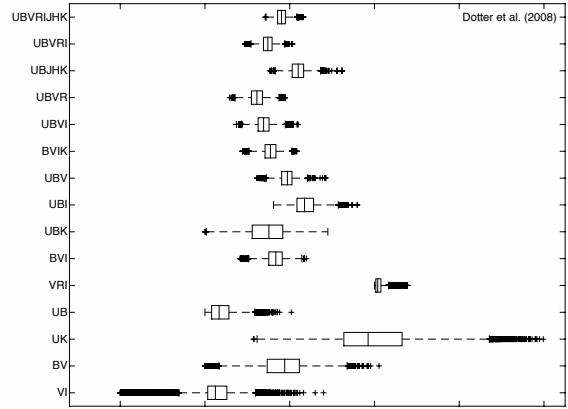}\\
    \includegraphics[angle=0,width=85mm]{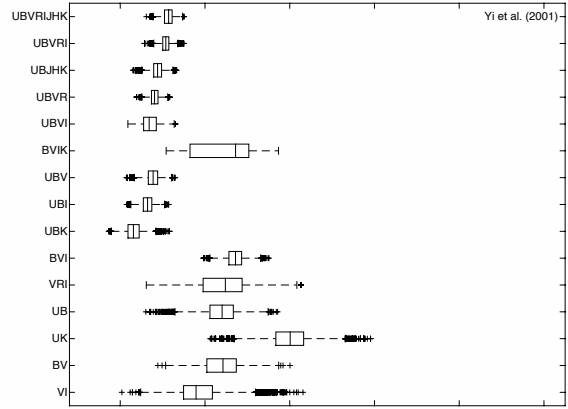}\\
    \includegraphics[angle=0,width=85mm]{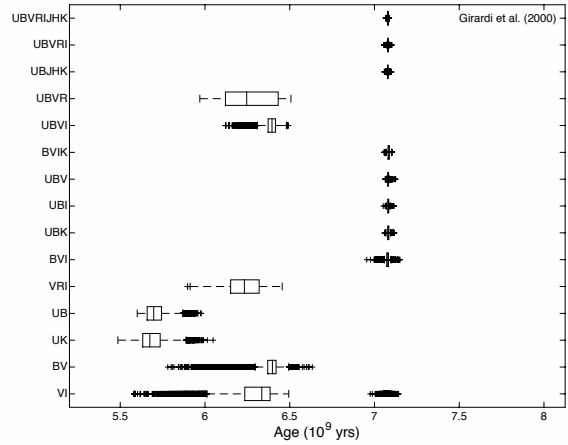}\\
  \end{tabular}
  \caption{Box-and-whisker plots for age across fifteen photometric band
	   combinations. Top: Dotter \etal (2008) isochrone models; Middle:
	   Yi \etal (2001) isochrone models; Bottom: Girardi \etal (2000)
	   isochrone models.}
  \label{fig:agebox}
\end{figure}

\begin{figure}[htp]
  \centering
  \begin{tabular}{cc}
    \includegraphics[angle=0,width=85mm]{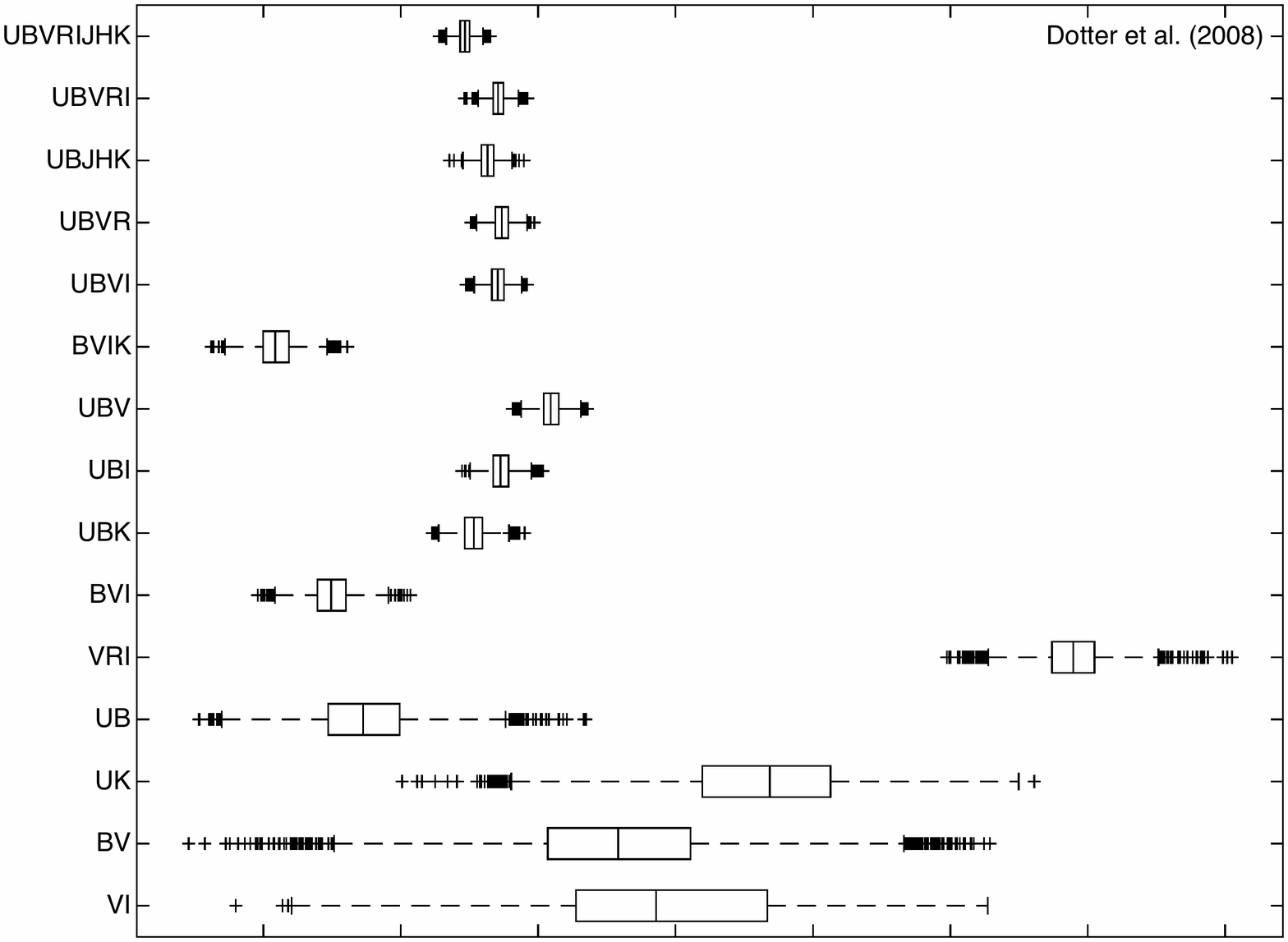}\\
    \includegraphics[angle=0,width=85mm]{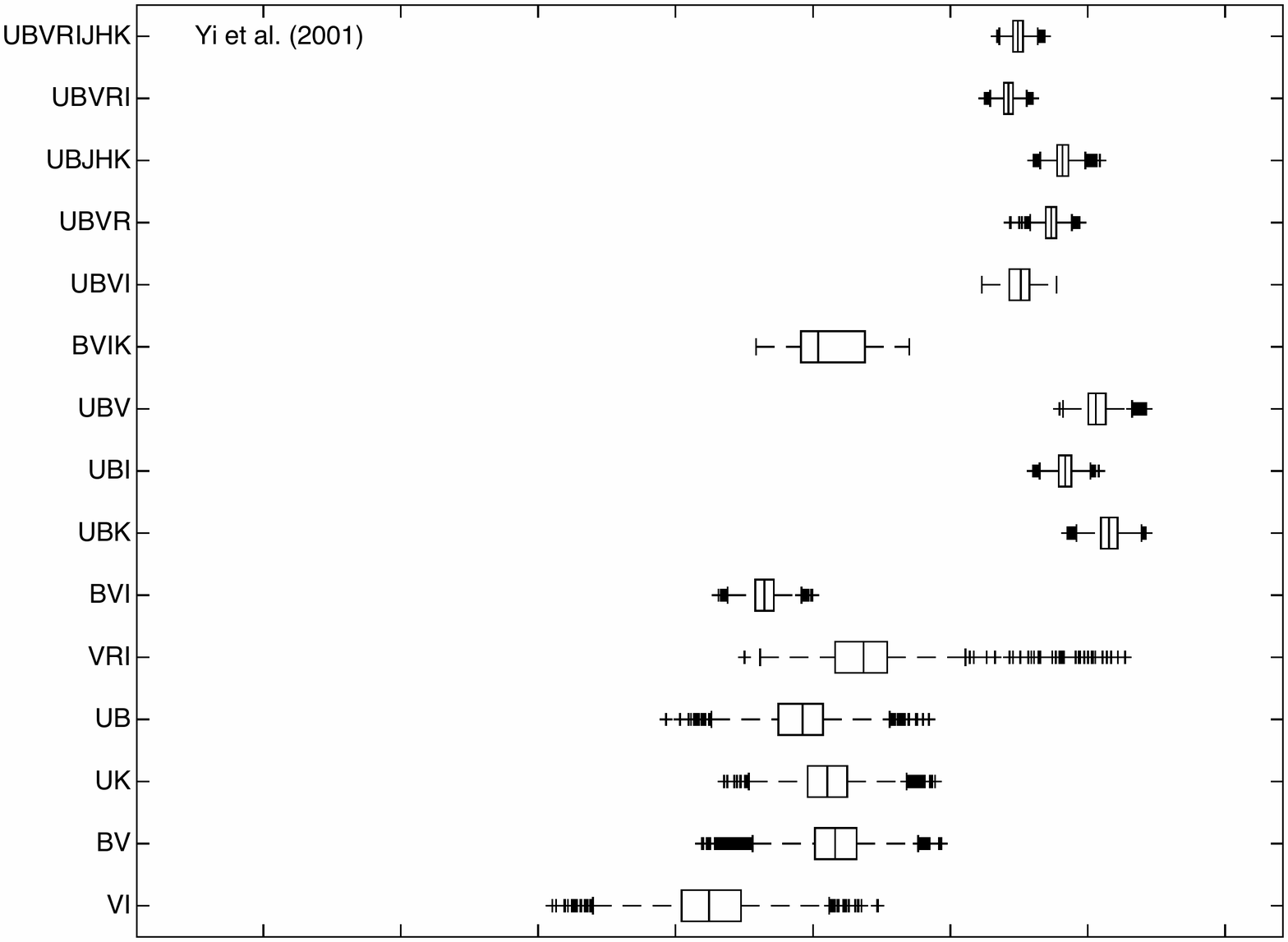}\\
    \includegraphics[angle=0,width=85mm]{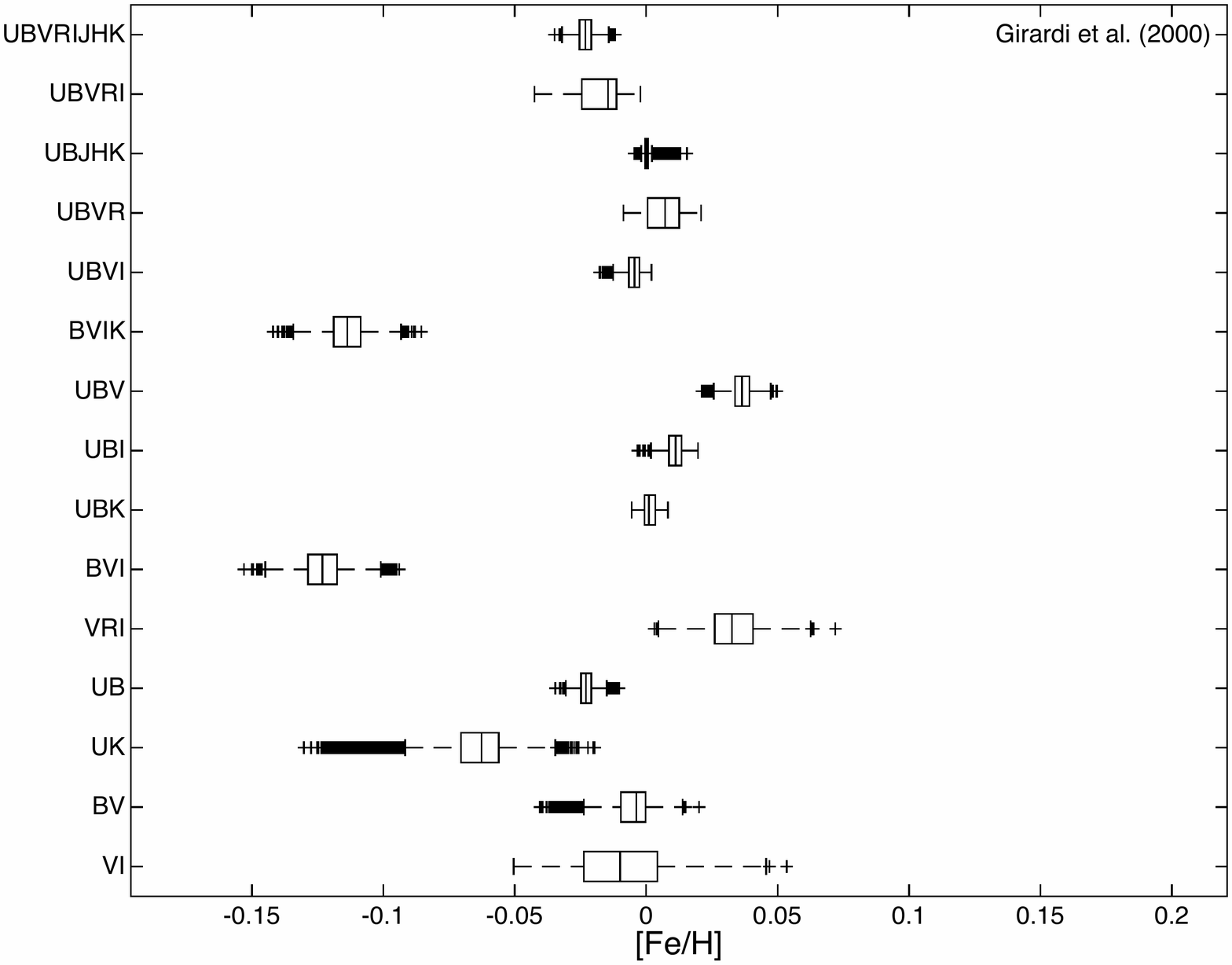}\\
  \end{tabular}
  \caption{Similar to \Fig{agebox}, but for [Fe/H].}
  \label{fig:metbox}
\end{figure}

\begin{figure}[htp]
  \centering
  \begin{tabular}{cc}
    \includegraphics[angle=0,width=85mm]{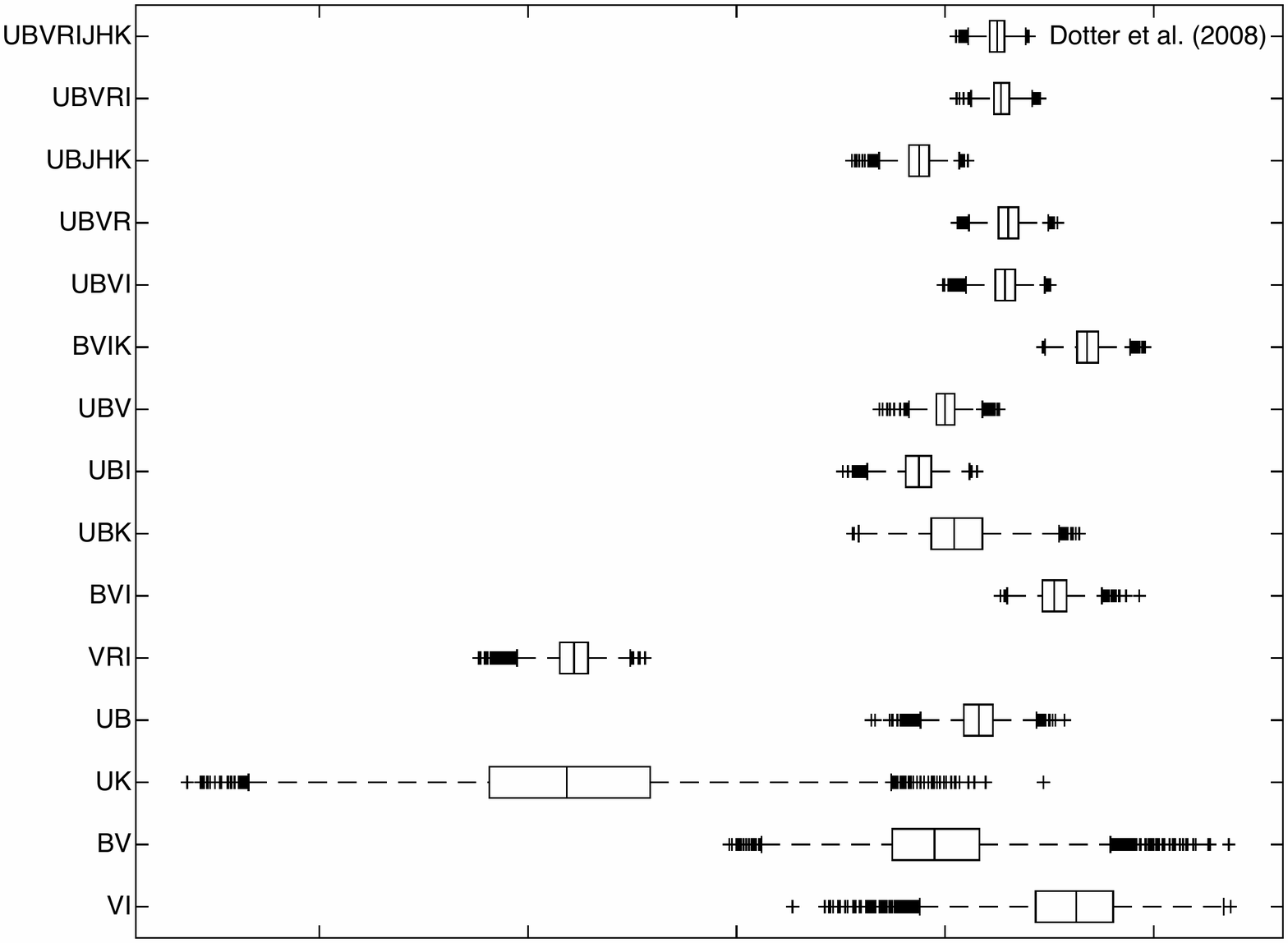}\\
    \includegraphics[angle=0,width=85mm]{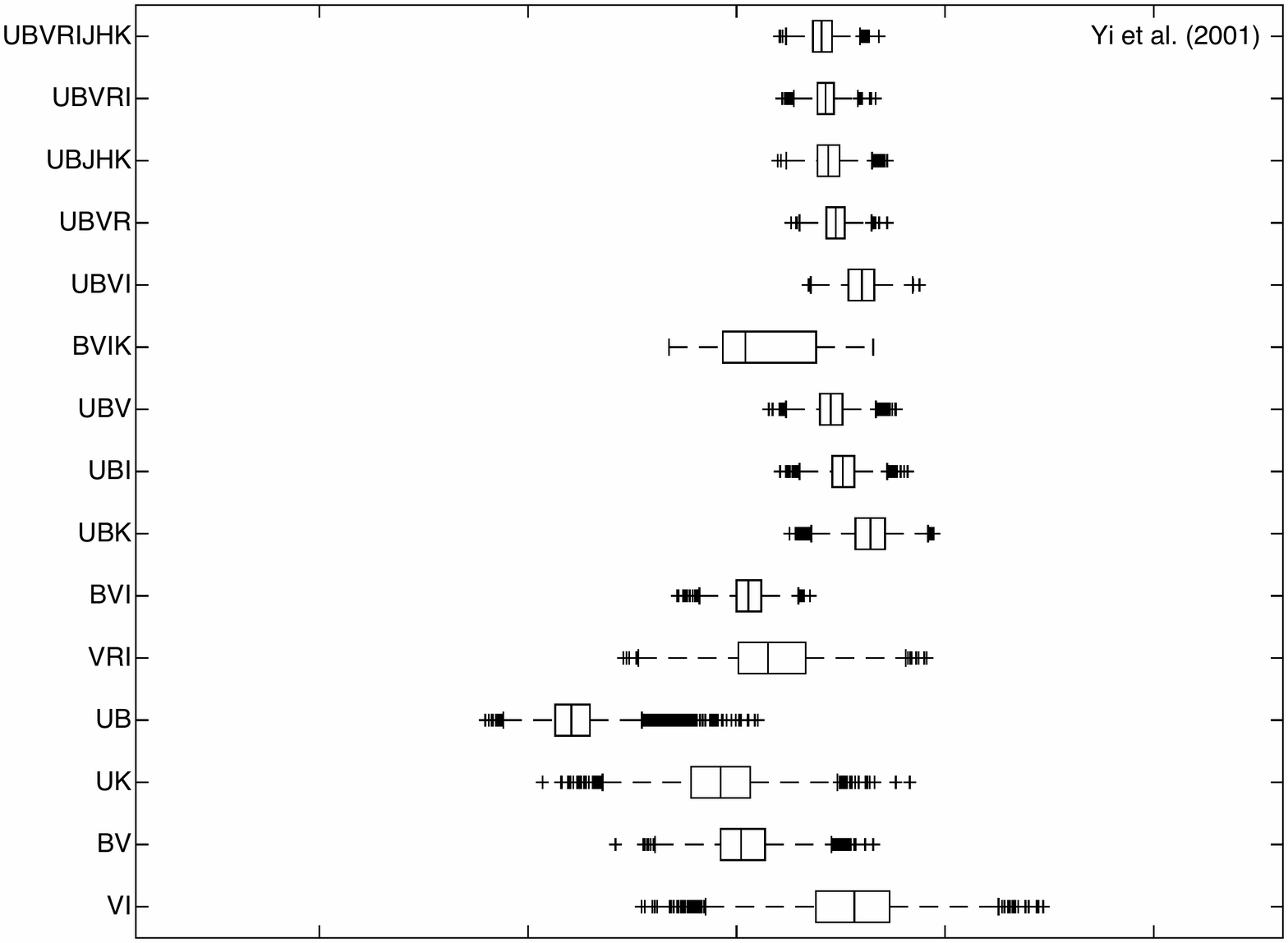}\\
    \includegraphics[angle=0,width=85mm]{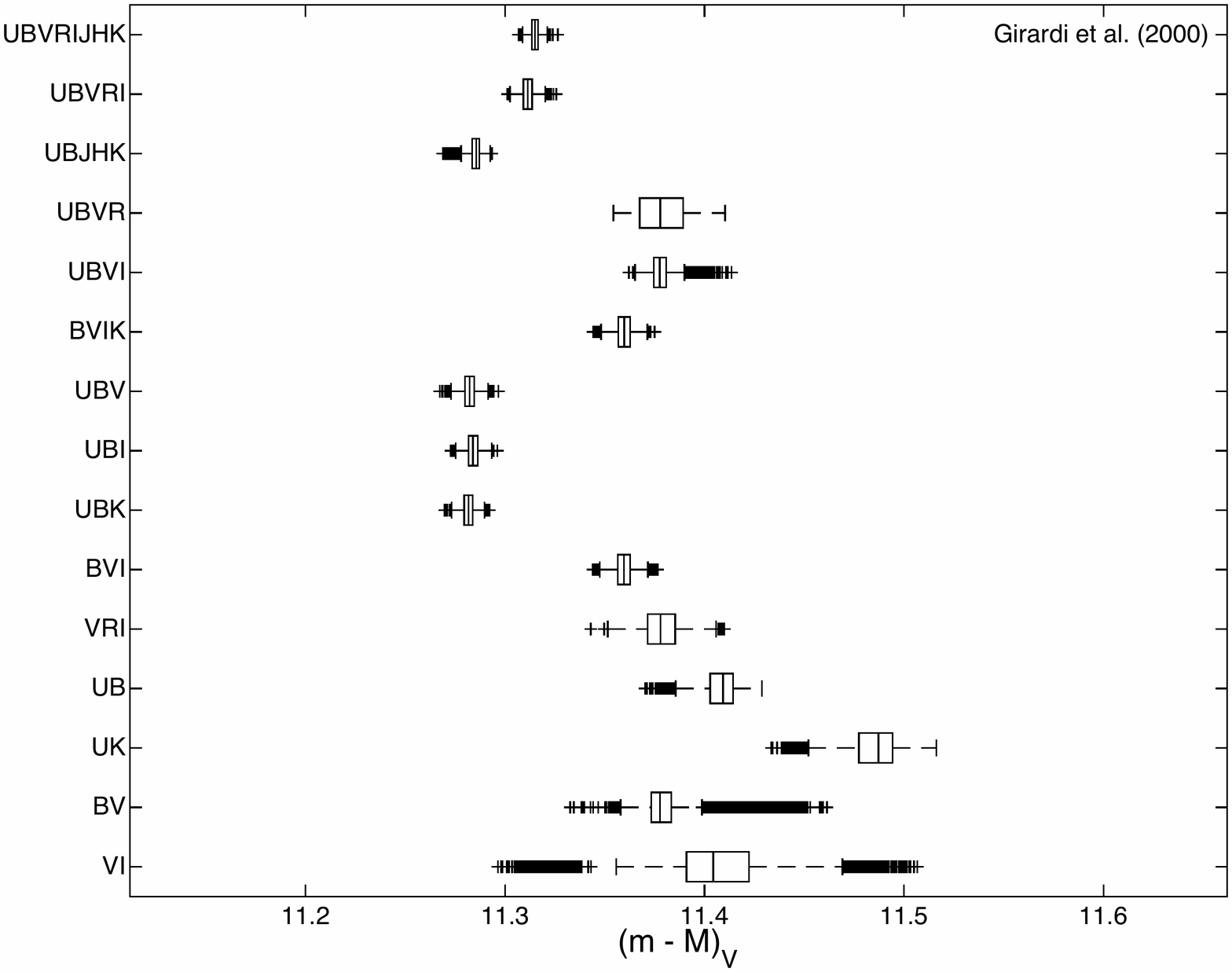}\\
  \end{tabular}
  \caption{Similar to \Fig{agebox}, but for $(m-M)_{\rm V}$.}
  \label{fig:modbox}
\end{figure}

\begin{figure}[htp]
  \centering
  \begin{tabular}{cc}
    \includegraphics[angle=0,width=85mm]{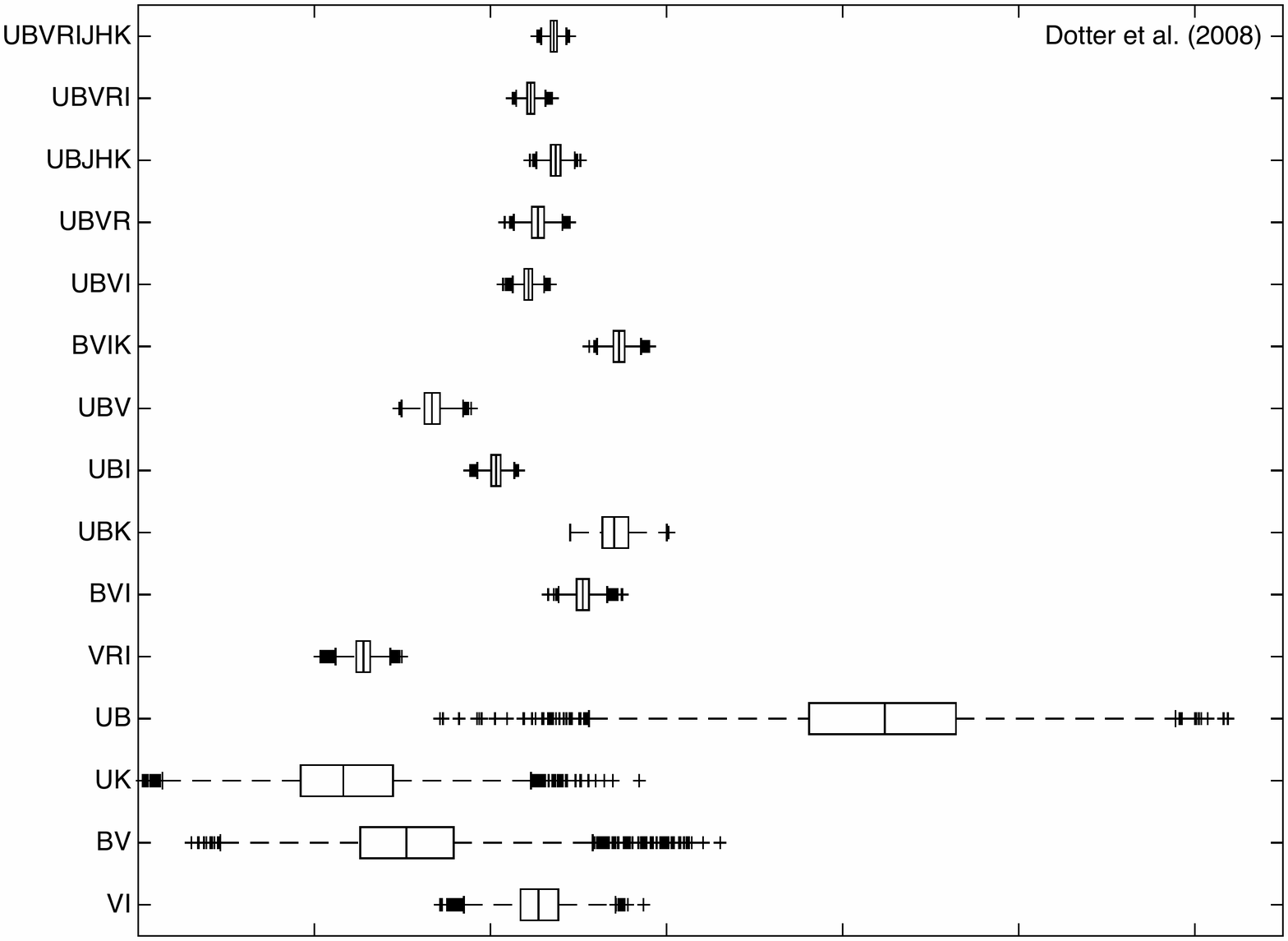}\\
    \includegraphics[angle=0,width=85mm]{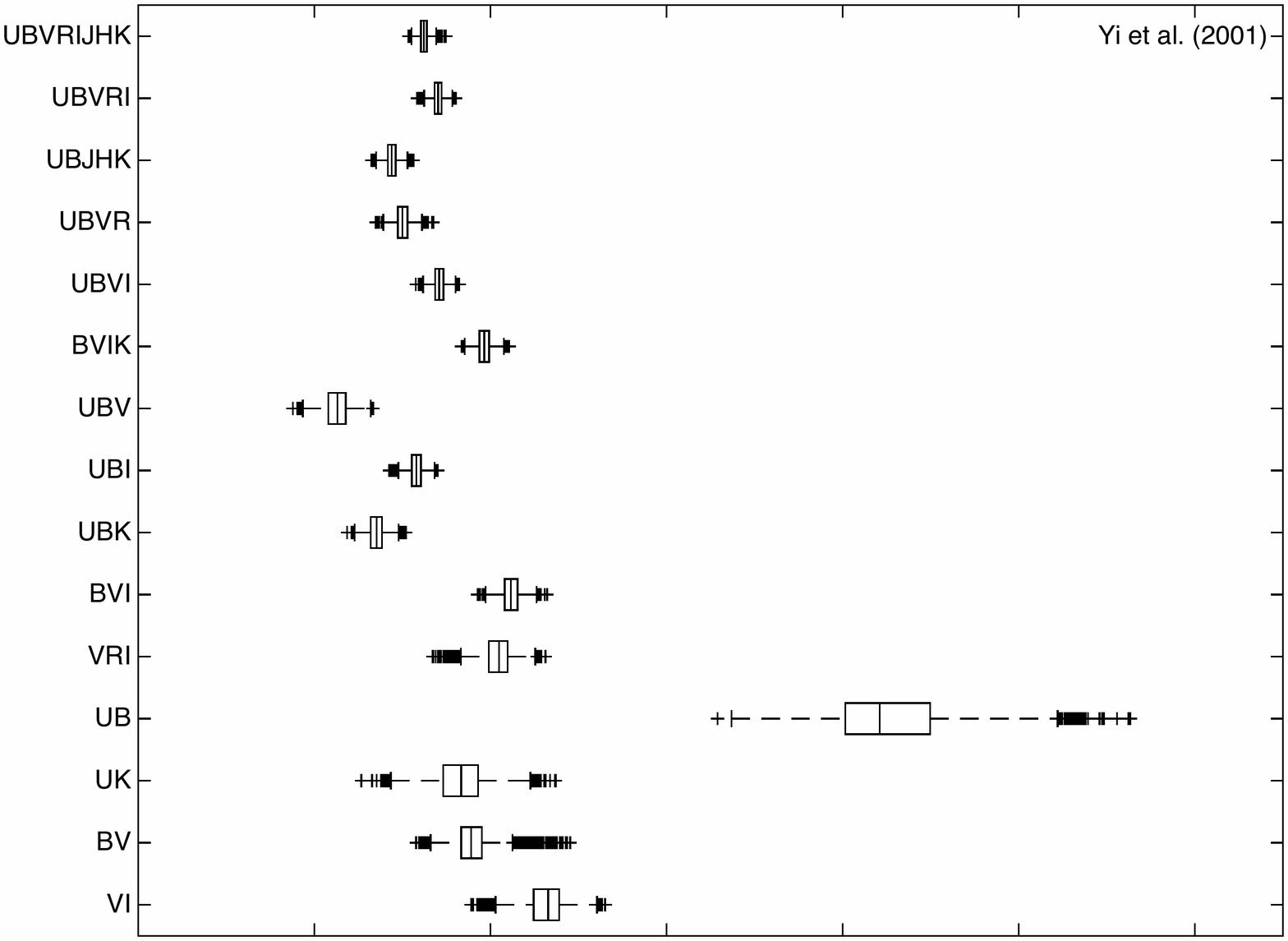}\\
    \includegraphics[angle=0,width=85mm]{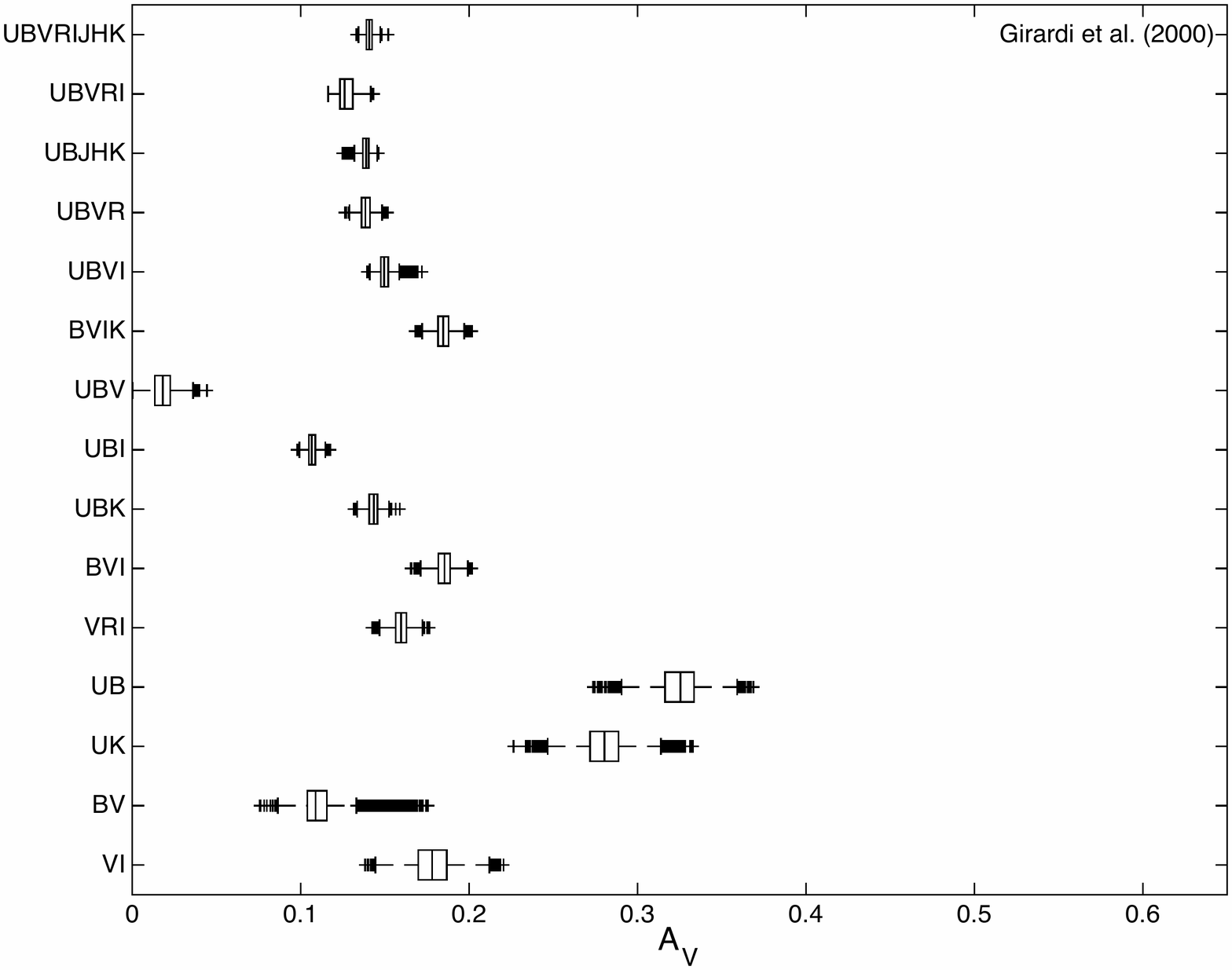}\\
  \end{tabular}
  \caption{Similar to \Fig{agebox}, but for $A_{\rm V}$.}
  \label{fig:absbox}
\end{figure}

In Figures \ref{fig:agebox} through \ref{fig:absbox}, we plot the derived
cluster parameters for NGC 188 for each of the three stellar evolution
models and each of the fifteen filter combinations of our study.  Focusing
first on \Fig{agebox}, we see that the Dotter \etal models converge to ages
that are internally consistent within the full range of the age posterior
distributions, except for $VRI$.  Fewer of these fits are consistent within
$\pm$ 1 $\sigma$, with $VRI$, $UB$, and $VI$ being the clearest examples.
Though the entire posterior distribution for $VI$ is wide, its systematic
offset from most of the other filter combinations is troubling given the
common use of this filter pair.  The age fits based on the Yi \etal results
show two age families that are again broadly consistent within the full
posterior ranges, though not within $\pm$ 1 $\sigma$ in many cases.  The
age fits based on the Girardi \etal models are displayed only for
completeness.  These models do not incorporate Equivalent Evolutionary
Points (EEPs, Bertelli \etal 1994), and so BASE-9 has difficulty
interpolating these models, particularly on the sub-giant branch and base
of the red giant branch.  This causes artificially narrow age locking in
about half of all cases.  These model fits are still useful, however, as
they show that even with a fixed (and reasonable) age, different filter
combinations may yield different values for the other cluster parameters
(see below), which further tests the reliability of fits as a function of
the filter combination that one employs.  Because of the EEP issue with the
Girardi \etal models, we do not report the fitted values based on these
models in the conclusions or abstract.

Which ages are most reliable?  Because stellar structure models are better
at predicting bolometric luminosity than $T_{\rm eff}$, which is not a
physical quantity in any case, and because stellar atmosphere models are
imperfect, we expect stellar evolution models to more poorly predict flux
in a particular passband than the sum of all available passbands.  The full
range of optical and near-IR filters from $U$ through $K_S$ does not
complete the stellar spectrum, of course, but for G and K stars, which
dominate NGC 188's CMD, the vast majority of the flux is within these
filters.  We therefore take the $UBVRIJHK_S$ fits as our reference standard
and expect that they will yield more accurate results than any other
combination of filters with less wavelength coverage.  We derive
significantly different ages from fitting Dotter \etal models (mean = 6.45,
median = 6.45, $q_1$ = 6.43, $q_3$ = 6.48, $\sigma$=0.04, all in Gyr) and
Yi \etal models (mean = 5.78, median = 5.79, $q_1$ = 5.76, $q_3$ = 5.81,
$\sigma$=0.03, all in Gyr).  \Fig{agebox} also shows that while the
distributions for some fits can be extremely narrow, with the central 50\%
of the distribution spanning less than 0.1 Gyr for a cluster $\sim$6 Gyr
old, or a precision better than 2\%, different age fits within a model set
can span nearly 1 Gyr in some cases, though typically differ by $\sim$0.4
Gyr.

\Fig{metbox} shows that the fitted metallicity distributions display
patterns similar to those of \Fig{agebox}, though the offsets in
metallicity are now so small that most are below the resolution limit of
current spectroscopic analyses.  In this case, the differences are
astrophysically unimportant, yet demonstrate that more filters tend to
yield tighter metallicity constraints and that some filter combinations can
provide statistically distinct fits at least for some models.  For example,
the $BVIK$, $BVI$, and $VRI$ Dotter \etal solutions differ significantly
from the other precise fits, as do two of these three combinations with the
other two stellar evolution models.  It is also evident that the $BVI$ fit
is not just the linear multiplication of the $BV$ and $VI$ fits.

\Fig{modbox} presents distance moduli fits for these model and filter
combinations.  In the full eight-filter cases, the fitted precisions are
excellent, with 50\% of the posterior distribution spanning $\leq$ 0.009
mag.  Yet, clearly, distances cannot be derived this precisely when the
distance moduli vary by 0.1 to 0.2 mag among different filter and
isochrone fits.

Finally, \Fig{absbox} presents absorption ($A_{\rm V}$) fits.  For NGC 188,
$A_{\rm V}$ is typically modest at $\sim$0.2 mag.  The $UBV$ and $UB$ fits
tend to be the least consistent with all the other filter combinations,
which is to be expected because the $U$ band is the most sensitive to
interstellar absorption.  On the other hand, it is surprising that the
$UBV$ and $UB$ fits are mutually inconsistent for all three isochrone sets
and that while $UB$ fits yield high absorption, $UBV$ appears to yield
overly low absorption.  Instead, these results more likely indicate that
models in $U$ and $B$ yield poorer fits to the data, likely due to the
narrow wavelength baseline and the difficulty of obtaining good stellar
atmospheres in the near-ultraviolet.

\section{Discussion}
\label{sec:discussion}

NGC 188 is one of the best-studied clusters on the sky.  It has modest
reddening, a well-constrained metallicity, and reliable cluster membership
probabilities.  In addition, this data set contains little photometric
contamination from binaries making it a good test case for the sensitivity
of isochrone fits to various filter combinations.  Figures \ref{fig:agebox}
through \ref{fig:absbox} demonstrate that different filter combinations can
yield meaningfully different ages and distances.  Specifically, we find
that the two wide-baseline filter combinations $UBVRIJHK_S$ and $UBVRI$
generally yield consistent results.  The infrared filters are thus less
essential.  There is somewhat weaker consistency between $UBVRIJHK_S$ and
$UBVRI$ on the one hand and $UBJHK_S$, $UBVR$, and $UBVI$ on the other
hand, or among these latter three combinations.  The other filter
combinations yield less consistent results, though these deviations may
often be lost in the widths of their substantially larger posterior
distributions.  Of particular historical note are the filter combinations
$UBV$, used extensively for photo-electric photometry, and $VRI$, used
extensively for CCD photometry.  In this study, we find that $UBV$ fits are
consistent with the eight-filter fits in age, somewhat less consistent in
metallicity and distance, and most inconsistent in absorption.  The $VRI$
fits are substantially less consistent with the eight-filter fits than the
$UBV$ fits for all four of these cluster parameters.  At this point we can
only affirm that these results apply to NGC 188 and probably other solar
metallicity clusters with similar ages.  We do not expect these same
disparity patterns for younger clusters whose bluer stars drive the age
fits and where bluer filters play a driving force in isochrone fitting.
Additionally, the consistency of fits may be improved for globular clusters
because low metallicity atmospheres are easier to model.

The inconsistencies inherent to isochrone fitting also highlight that
stellar models still suffer from incomplete treatment of important physics.
Because any stellar evolution model has a specific relationship between
turn-off mass and age, and because mass is precisely connected to
luminosity, the other three cluster parameters can act as free parameters,
at least within some constrained bounds.  To match the luminosity of the
turn-off, BASE-9 can adjust the distance with appropriate additional
adjustments in absorption and metallicity.  The freedom of adjusting
multiple cluster parameters may bury the evidence that would be most
helpful in determining which stellar models and which wavelength ranges are
most problematic.  In principle, highly constrained isochrone fits for a
wide range of clusters with different ages and metallicities may reveal the
magnitude of the underlying physical problems, whether they be assumptions
about convection at the base of the giant branch or line blanketing in
stellar atmospheres or perhaps other physics that we may be less concerned
about.

In any case, the European Space Agency's GAIA satellite mission heralds
a new era where at least one large source of uncertainty, cluster distances,
can be highly constrained with great precision and accuracy.  Locking down
cluster distances with reliable parallaxes to 24 $\mu$as (expected for a
$V \leq 15$ star with G star colors, see
http://sci.esa.int/gaia/47354-fact-sheet), could constrain the distance to
NGC 188 (at $\sim$2 kpc) considerably.  There are 27 stars in our CMD with
$V \leq 15$ and 83 stars with $V \leq 15.5$.  The expected parallax
accuracy of 4.8\% per star at 2 kpc improves by at least a factor of
$\sqrt{27-1} \approx 5$, and could be more than 10 times higher once all
observed stars are properly included, meaning that the distance to NGC 188
will be known to at least 0.5-1\%.  This corresponds to an uncertainty of
0.01-0.02 mag in distance modulus, which is more than an order of magnitude
improvement over the range of the fits in \Fig{modbox}.

\section{Conclusions}
\label{sec:conclusion}

Using the Bayesian statistical stellar evolution package BASE-9, we fit the
well-studied old open cluster NGC 188 for age, [Fe/H], $(m-M)_{\rm V}$, and
$A_{\rm V}$ under fifteen different photometry regimes, using a range of
filters and wavelength baselines.  We argue that employing all eight
filters, and thus the widest baseline, yields the most precise and accurate
fits.  The five-filter $UBVRI$ combination was nearly as good.  However,
other filter combinations often gave inconsistent results with each other
and with the eight-filter results.  These inconsistencies can span 1 Gyr,
though 0.4 Gyr differences, or $\sim$6\%, are more typical for NGC 188.

Differences amongst the model sets can also be substantial.
Specifically, fitting Yi \etal (2001) and Dotter \etal (2008) models
to the eight-filter data, yields the following mean cluster parameters:
age = \{5.78 $\pm$ 0.03, 6.45 $\pm$ 0.04\} Gyr, 
[Fe/H] = \{+0.125 $\pm$ 0.003, $-$0.077 $\pm$ 0.003\} dex,
$(m-M)_{\rm V}$ = \{11.441 $\pm$ 0.007, 11.525 $\pm$ 0.005\} mag, and
$A_{\rm V}$ = \{0.162 $\pm$ 0.003, 0.236 $\pm$ 0.003\} mag,
respectively.  With such small formal fitting errors, these two fits
are substantially and statistically different.  The differences amongst
fitted parameters using different filters and models is a cautionary tale
regarding our current ability to fit star cluster CMDs.

Our case study of NGC 188 should be extended to other stellar clusters
to cover the widest range of cluster parameters.  In doing so, the match
between the filter response functions for the theoretical evolutionary models
and the actual observations must be confirmed to eliminate a common source
of systematic error.

\acknowledgements

We thank Elliot Robinson for his help with BASE-9 development.  This
material is based upon work supported by the National Aeronautics and Space
Administration under Grant NNX11AF34G issued through the Office of Space
Science.  SC acknowledges support through a Discovery grant from the Natural
Sciences and Engineering Research Council of Canada.  AMG is funded by
a National Science Foundation Astronomy and Astrophysics Postdoctoral
Fellowship under Award No. AST-1302765.

\end{document}